# CHARACTERISATION OF THE ETCHING QUALITY IN MICRO-ELECTRO-MECHANICAL SYSTEMS BY THERMAL TRANSIENT METHODOLOGY

*Péter Szabó[1,2], Balázs Németh[1], Márta Rencz[1,2], Bernard Courtois[3]*

[1]Department of Electron Devices, Budapest University of Technology & Economics,
H-1111 Budapest, Goldmann Gy. tér 3., Hungary

[2]MicReD Ltd., Budapest XI, Etele u. 59-61, H-1119 Hungary

[3]TIMA Lab. Grenoble, France

<szabop | rencz>@micred.com

<szabo | nemethb>@eet.bme.hu, Bernard.Courtois@imag.fr

## ABSTRACT

Our paper presents a non-destructive thermal transient measurement method that is able to reveal differences even in the micron size range of MEMS structures. Devices of the same design can have differences in their sacrificial layers as consequence of the differences in their manufacturing processes e.g. different etching times. We have made simulations examining how the etching quality reflects in the thermal behaviour of MEMS structures. These simulations predicted change in the thermal behaviour of MEMS structures having differences in their sacrificial layers. The theory was tested with measurements of similar MEMS devices prepared with different etching times. In the measurements we used the T3Ster thermal transient tester equipment. The results show that deviations in the devices, as consequence of the different etching times, result in different temperature elevations and manifest also in shift in time in the relevant temperature transient curves.

## 1. INTRODUCTION: THERMAL TRANSIENT TESTING OF THE ETCHED LAYER OF MEMS

MEMS devices are typically used in areas where low power consumption, small size and special abilities are required. The proper functioning of a MEMS device is strongly dependent on the quality of the manufacturing processes. On the other hand the reliability is very important in most of the applications they are used.

The etching process is one of those that can be responsible for the malfunctioning of MEMS devices, and the control of it is not easy. In this manufacturing process step portions of different deposited sacrificial layers are removed by fluids in case of wet etching or ionised gas in case of dry etching. The quality of etching in the wet etching process is strongly dependent on the material to be etched, on the concentration of the etching chemical and on the etching time. In dry etching the pressure, the etching gas and the etching time are the main points of the settings. If an etched layer contains parts remained in the etched layer then moving parts of the device may be immobilised. On the other hand, over-etching the remaining parts in non-sacrificial layers can cause fractures and sticking. To check the quality of the etching is a crucial point in the manufacturing. There are several ways that can be used to investigate the state of the device: e.g. Optical and, or Scanning Electron Microscopy, or to check the functioning of the device. These techniques are applicable only in finding mistakes on the surface or in identifying elements not shrouded by others. Unfortunately to get information from deeper levels they are barely usable.

We found that the thermal transient method [1] can be used in several MEMS devices to discover the etching problems. This analysis method and the T3Ster equipment [2], applying this, have been investigated and evaluated at the Department of Electron Devices of BUTE and MicReD. The method and the equipment were developed originally for the measurement of thermal resistances of semiconductor devices, but the equipment has a broad scope of other use as well. It can measure the thermal behaviour of a variety of elements showing change in their electrical parameters by temperature change. MEMS devices usually have elements that have temperature dependent electric parameters e.g. electric resistances of bridges, thermopiles etc.

In the transient measurement the electrical resistance of the MEMS element is driven with a certain current. After the temperature stabilisation the current is switched off and the voltage change on the electric resistance induced by the temperature change is measured. The maximal excursion of the curve gives the temperature change as a result of the applied power change. From the temperature elevation and the length of the transient in time the etching ratio between the different samples can be determined. In the subsequent sections we show the method through simulations and measurements of MEMS devices.

In order to carry out experiments we needed special MEMS samples. These samples are of the same design, they went through the same manufacturing steps, except the etching. The etching times were different for the different dies we have examined. The samples have been manufactured for the experiments at the TIMA laboratory [3].

## 2. THE STRUCTURE UNDER EXAMINATION

Three chips containing MEMS devices on their surfaces were fabricated with different etching times. The chips had not only the same structures but were packaged also similarly. Figure 1 shows the surface photo of the investigated chip with MEMS devices.

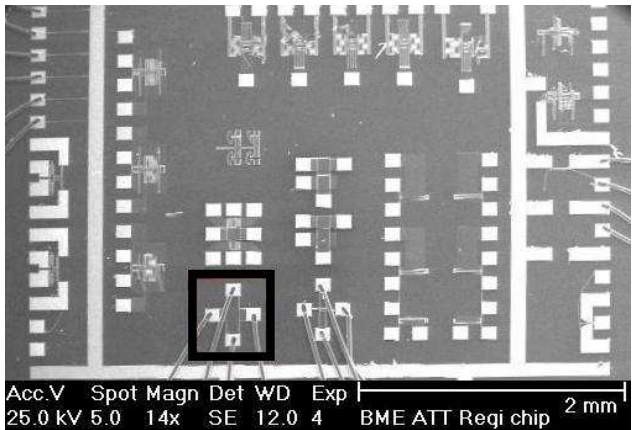

**Figure 1  SEM picture of the chip, the investigated resonator in the black frame**

Before starting the measurement we checked all the elements with optical and scanning electron microscopy to reveal the unusable structures, as the etching process destroyed many elements. An electro mechanic resonator seemed to be the most appropriate for the measurement, it is shown in the black frame in Figure 1. Its larger view is presented in Figure 2.

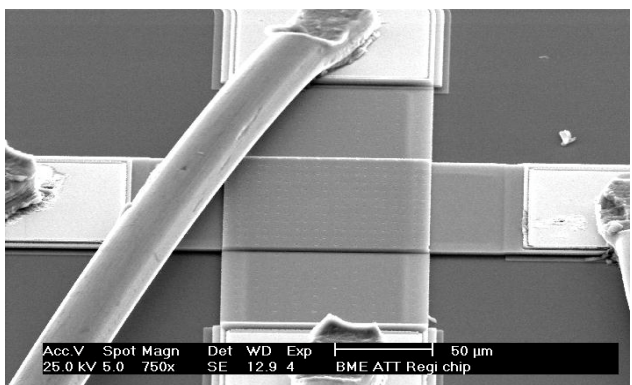

**Figure 2  The investigated electro mechanic resonator**

The chip was fabricated by PolyMUMPs technology [4]. The cross sectional view of the layers and the constituting materials is shown in Figure 3. The two resonator beams were fabricated in the polysilicon layers by etching off the

PSG layers between them. The property of the etched PSG layers play the most important role in the functioning of this device, as the resonance ability is strongly dependent on the remained parts of the PSG layers.

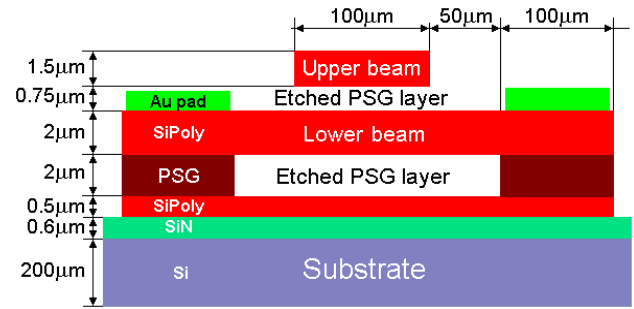

**Figure 3  Cross sectional view of the resonator**

The ends of the beams contain gold pads which were bonded directly to the pins of the package.

## 3. THERMAL SIMULATIONS OF THE STRUCTURE

We carried out simulations to examine the thermal properties of the structure using the SUNRED simulator [5]. We were interested in the effect of the etching quality on the temperature distribution on the surface and on the shape of the thermal transient curves. The simulations were made on the quarter of the structure as the structure shows spatial symmetry. The first model describes the fully etched state, that is, when all the PSG between the layers is removed. In the first simulation the power was applied to the upper layer. The steady state temperature distribution for this case is shown in Figure 4. The dimensions on the structure are not to scale to enable better visibility of the results

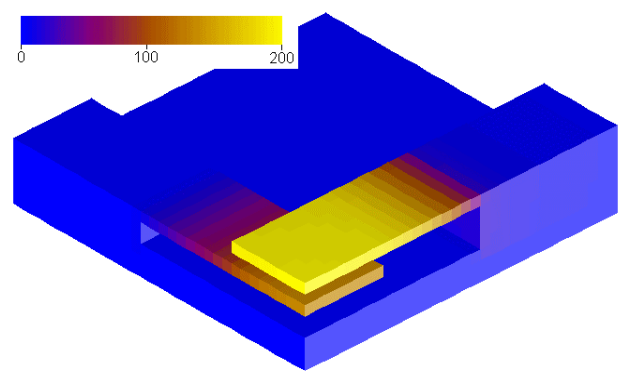

**Figure 4  3D Temperature distribution of the resonator, the upper layer dissipates, the PSG layers are fully etched, equidistant mesh**

In the second simulation we drove the lower layer. The temperature distribution for this case is shown in Figure 5.

It can be seen that only the suspended layers have elevated temperature, and mainly in their middle, the ends are practically cold in both simulations. The reason of this is that the silicon has good thermal conductivity and conveys the heat away easily towards the ambient.

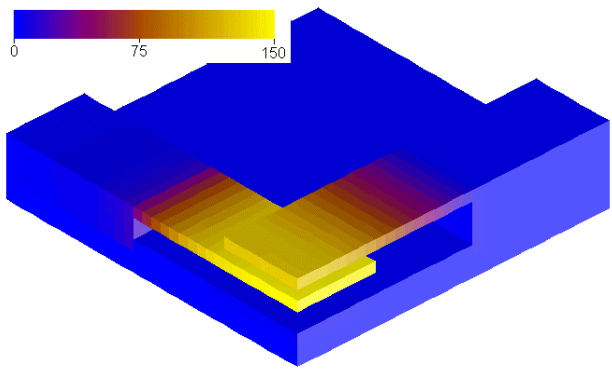

**Figure 5  Temperature distribution of the resonator, fully etched, the lower layer dissipates, the PSG layers are fully etched, equidistant mesh**

We can also see in Figure 5 that the temperature elevation in the second simulation is smaller than in the lower layer driving case. Consequently, for higher effect it is better to drive the upper layer.

We simulated the effect of partial etching in an other model. This was created by leaving one tenth of the PSG between the layers and the silicon substrate in the region where the beams are overlapped. In Figure 6 the effect of this action can be examined on the temperature distribution.

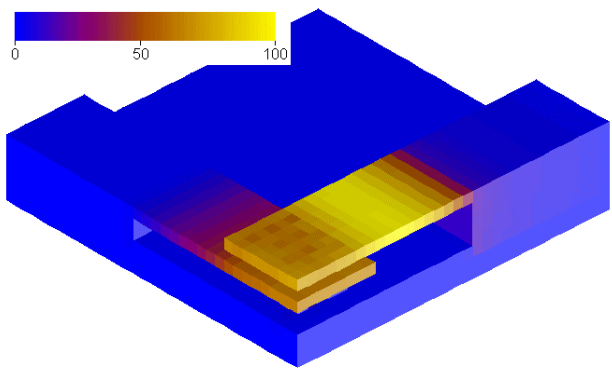

**Figure 6  Temperature distribution of the resonator, the tenth of the PSG is remained between the two layers, the upper beam dissipates, equidistant mesh**

It is observable in the figure that the hot part of the temperature distribution is shifted towards to the ends of the upper beam. Where the layers are overlaping the temperature is lower, as the consequence of the lower thermal resistances towards the chip. As it can be seen the temperature distribution is not homogeneous in the overlapping beams as a result of the inhomogeneous material distribution of the remained parts of the PSG. This simulation showed that even a very low amount of material (1/10 of PSG) in the etched layer that is in connection with the upper and lower layers can cause very large differences in the temperature. The inhomogenity of the temperature distribution may cause inhomogeneous stresses in the material and be responsible for cracking. We simulated also the thermal transients on the fully and the partially etched structures [6]. The upper beam was powered in case of each simulated curve, and the temperature was measured in the middle of the beam. The results can be seen in Figure 7. The curve with the larger temperature excursion corresponds to the fully etched model and the curve with the lower excursion refers to the partially etched structure.

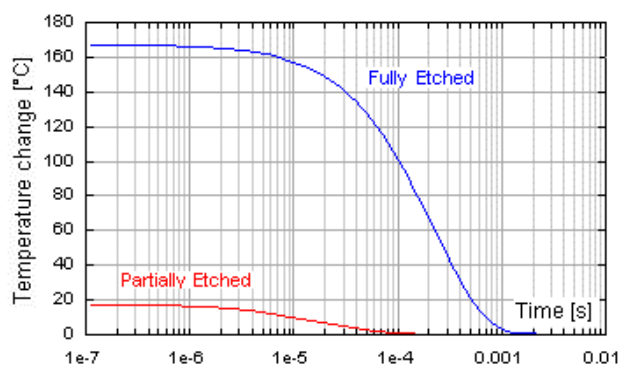

**Figure 7  Simulated thermal transient response of the upper beam, model of fully and partially etching**

As it was mentioned above the transient curve shifts towards the left, if the etched layers have remained parts. This effect can be well observed in Figure 8. For the sake of better comparison we multiplied the temperature values of the bad etching model by 9.6 to have the same temperature excursion.

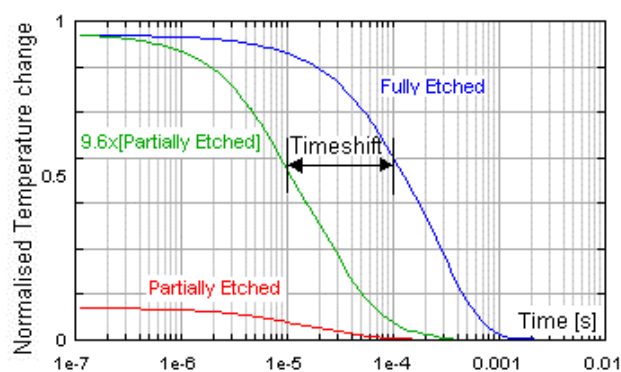

**Figure 8 The shift of the temperature transient curve in time with diminishing thermal resistance**

It can be seen that the shift in time is about one decade while the change in temperature is about one tenth as well. This is not accidental, it comes from the principle of the time constant. The time constant is the coefficient of the exponential relaxation of the temperature at the point (region) where the power step is applied. Each thermal system has a characteristic (average) thermal resistance $R_{th}$ and thermal capacitance $C_{th}$, the multiplication of these gives the value of the time constant:

$$\tau = R_{th} * C_{th} \qquad (1)$$

If a system has a one tenth of thermal resistance then the time constant is also the one tenth. This principle results in a nearly the same factor change (1/10) in the temperature elevation and the time shifting in the thermal transient curves referring to the two examined cases.

## 4. CALIBRATION OF THE TEST STRUCTURES

In our experiments we measured three samples. Each was etched for different length of time: 30s, 60s and 90s. For the temperature measurement we used the temperature induced change of the electric resistances of the beams. The change of the electric resistance is described with the linearized equation of:

$$\Delta R_{el} = \alpha * R_{el0} * \Delta T \qquad (2)$$

where $\alpha$ is the coefficient, $R_{el0}$ the electric resistance at a given $T_0$ temperature, $T$ the actual temperature, $\Delta T$ the temperature difference ($\Delta T = T_0 - T$) and $\Delta R_{el}$ is the change of the electric resistance. In the calibration process we have put the device onto a cold-plate, and drove it with a 25mA sensor current applied at each beam simultaneously. We changed the temperature of the cold-plate and captured the voltage change along the temperature change. The circuit of this setup is shown in Figure 9.

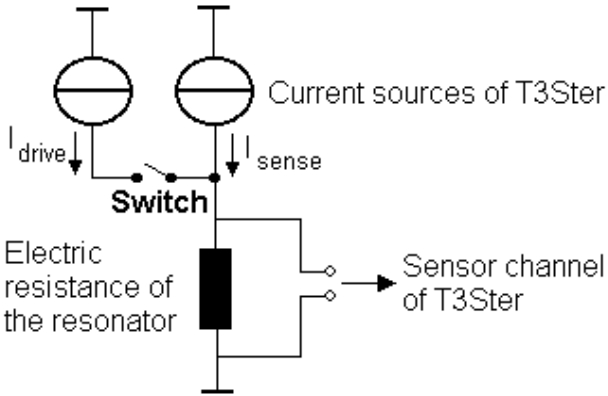

**Figure 9  The electronic connection of the calibration and measurement setups**

The sensor sources and voltage measurement channels of the T3Ster thermal transient tester were used in the calibration [1], [2]. The voltage change results for the upper beams of the three samples are presented Figure 10. The curves show good linearity, consequently we can use the linearised formula and the $\alpha$ constant can be easily calculated.

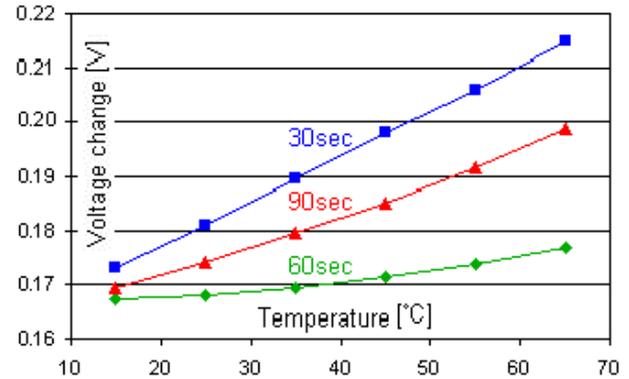

**Figure 10  Calibration curves of the upper beams**

## 5. TRANSIENT MEASUREMENTS

During the thermal transient measurement we drive the device continuously with a sensor current in this measurement, we used $I_{sense}$=25mA, and with a driving current, in this measurement $I_{drive}$=25mA, see, Figure 9. We wait until the system reaches the thermal steady state. After this we switch off the driving current in $\mu$s time and capture the voltage transient with $\mu$s sampling rate in logarithmic sampling mode (200 sample/octave). The temperature transient curve can be calculated from the electric transient, using the $\alpha$ constant obtained in the calibration process.  Usually immediately after switching, the first part of the thermal transient curve is influenced by the parallel electrical transients occurring in the electrical system, e.g. discharging of capacitances, etc. To examine the strength of this effect we made measurements with simple, commercial resistors having the same resistances as the beams of the examined structure. The model system consists of a 100$\Omega$ modelling the upper beam and a 33$\Omega$ modelling the lower beam. We measured the electrical transient at the 100$\Omega$ with 25mA sensor and 25mA driving currents, Figure 11.

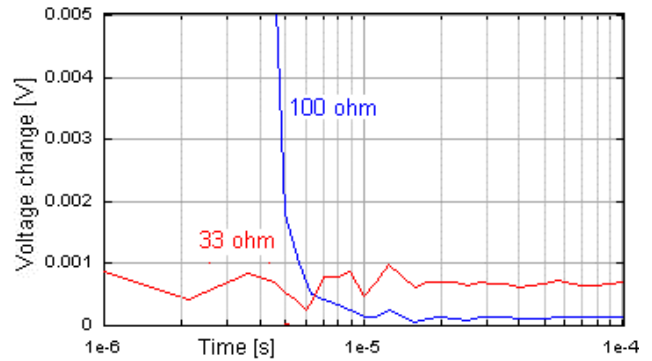

**Figure 11  Smoothed electric transients in the model system,  25mA sensor and 25mA driving current**

In the measurement we also measured the voltage change of the 33 Ω (with 25mA sensor current) to check the coupling electric cross effects between the measured channels. Note that all transients ended till 5-10μs only some noise can be seen after this point.

We wanted to know the parasitic effect of the electric capacitances existing in the examined MEMS device. The calculated parasitic electric capacitance of the device is under 20pF. According to this data we made measurement in the model system with the same setup as before, but connected an overestimated 82-pF capacitance parallel with the 100Ω. The electric transient result is shown in Figure 12. This curve is very similar to the transient in Figure 11. We do not show the curve of the voltage change of the 33Ω, as it is same as in the previous measurement, see Figure 11.

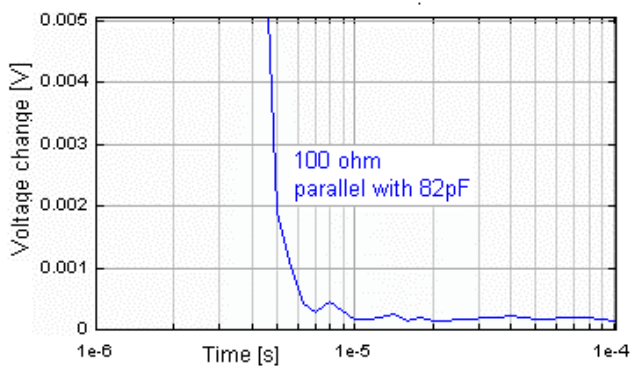

**Figure 12  Effect of the 82 pF connected parallel with the 100 Ω, in the model system**

In the next step we measured the MEMS device that was etched for 60s. Each beam was connected to a sensor channel, but none of them was connected to a sensor current source. We drove the upper beam with 25mA, and made the transient measurement, see Figure 13.

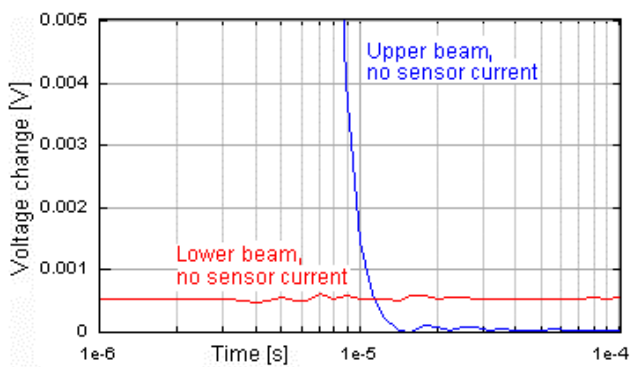

**Figure 13  Electric transients of the beams of the electrostatic resonator, no sensor current was applied at the beams, the upper beam was driven with 25mA**

We can see that there is practically no change after 15μs. The curve shifted to the right with 5μs compared to the model system Figure 12, Figure 13.

In the next measurement we applied 25mA sensor current at the lower beam. The upper beam had no sensor current, but it was powered with a driving current. A thermal transient measurement was accomplished again. Now we experienced the appearance of a bump in the voltage transient of the lower beam, Figure 14.

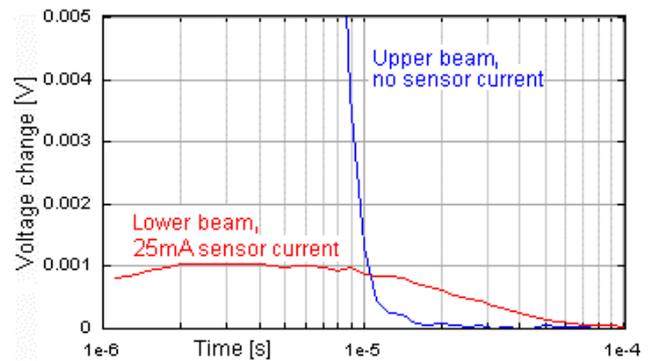

**Figure 14 Electric transients of the beams of the 60s etched electrostatic resonator, no sensor current was applied at the upper beam, lower beam had 25mA sensor current, the upper beam was driven with 25mA**

The transient of the upper beam is the same as before, see Figure 13, and Figure 14. The bump in the transient of the lower beam has to be the effect of the electric resistance change induced by the temperature change. It is not a parasitic effect as it was not present in the previous measurement in Figure 13. It was not present in the measurement of the model system, Figure 11, so it is neither the parasitic effect of the measurement itself, nor that of the system. In the next step we applied sensor current to the upper beam and did not apply it to the lower beam. The transient results are shown in Figure 15.

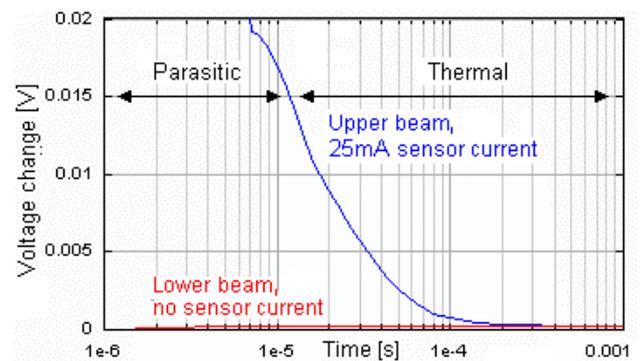

**Figure 15 Electric transients of the beams of the electrostatic resonator, sensor current was not applied at the lower beam, upper beam had 25mA sensor current, the upper beam was driven with 25mA**

At the lower beam we can see practically no change, as it had no sensor current. We can see a large bump in the transient of the upper beam. The bump begins at about 8$\mu s$ and holds practically till 200$\mu s$. The bump in the transient of the upper beam appeared when we applied the sensor current The bump of the transient of the lower beam has disappeared as no sensor current was applied. These experiments demonstrate that what we observe is the result of a solely thermal effect. This means that the method is in fact applicable to detect the presence of improper etching

After these preliminary measurements we checked the effect of the etching time on the thermal transient curves. We measured the transients of the 60 and 90 sec etched samples driving the upper beam. We transformed the voltage transients to temperature transients with the sensitivities derived from the calibration, the results are shown in Figure 16. We can see that the temperature elevation of the sample etched for 90s is about four times higher than that of the sample etched 60s. We also experience a time shifting of the transient of the device etched for 90s. For better visibility of the time shift we multiplied the curve by 4. The temperature elevation can be read at about 10$\mu s$ as the parasitic electric transients end at that point. At this point in time the sample etched for 60s has about 50K and the sample etched for 90s has about 300K temperature elevation.

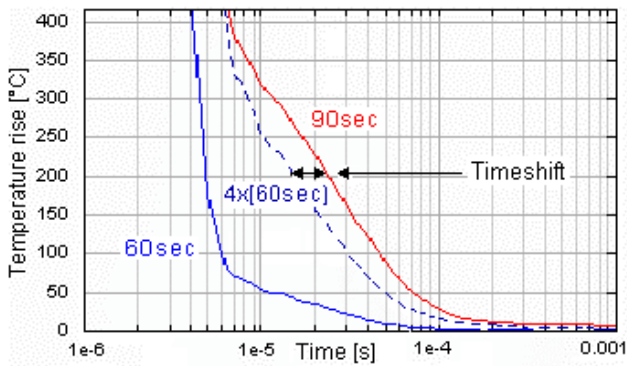

**Figure 16  Temperature transient of the upper beam of the electrostatic resonators etched at different times, the upper beam was driven with 25mA**

The measured result of the whole temperature elevation of the device etched for 90s correlates well with the result of the simulation of the fully etched upper beam driven model presented in Figure 4.

## 6. CONCLUSION

We have presented a non-destructive thermal transient measurement method that can be applied to discover etching problems in MEMS devices. Experiments have shown that remained material in the etched layer of a MEMS device can drastically change the thermal properties of that device. We have demonstrated that thermal transient measurement can be used to reveal etching problems if there is a resistor in the structure that can be contacted electronically. The feasibility of the method was presented in the examination of the etching quality of electro-mechanic resonator structures. With the help of the thermal transient measurements the samples with insufficient etching could be identified from the lot of samples containing both good and insufficiently etched devices.
We will continue our work in this field by checking the sensitivity of the method on a variety of devices.

**Acknowledgements**

This work was supported by the PATENT IST-2002-507255 Project of the EU and by the OTKA-TS049893 project of the Hungarian Government.

## 7. REFERENCES

[1] V. Székely and Tran Van Bien: "Fine structure of heat flow path in semiconductor devices: a measurement and identification method", Solid-State Electronics, V.31, pp. 1363-1368 (1988)
[2] www.micred.com/t3ster.html
[3] http://tima.imag.fr/
[4] www.memsrus.com/nc-mumps.poly.html
[5] V. Székely: SUNRED: "A new thermal simulator and typical applications", Proceedings of the 3rd THERMINIC Workshop, September 21-23, 1997, Cannes, France, pp. 84-90.
[6] P. Szabó, O. Steffens, G. Farkas, Gy. Bognár: "Thermal transient characterization methodology for packaged semiconductors and MEMS structures", Proceedings of the 12[th] MIXDES Conference, 22-25 June 2005, Krakow, Poland, pp. 271-276